# Photo-active collagen systems with controlled triple helix architecture

Giuseppe Tronci,[1,2] Stephen J. Russell,[2] David J. Wood[1*]

[1] Biomaterials and Tissue Engineering Research Group, Leeds Dental Institute, University of Leeds, UK

[2] Nonwovens Research Group, Centre for Technical Textiles, University of Leeds, UK

**Table of contents entry**

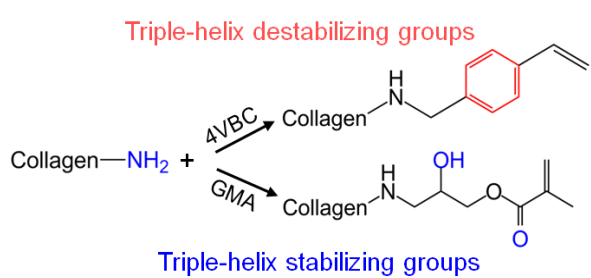

Covalent functionalization of type I collagen with photo-active moieties of varied backbone rigidity resulted in triple helical systems of varied network architecture, so that bespoke structure-property relationships could be established.

**Abstract**

The design of photo-active collagen systems is presented as a basis for establishing biomimetic materials with varied network architecture and programmable macroscopic properties. Following in-house isolation of type I collagen, reaction with vinyl-bearing compounds of varied backbone rigidity, i.e. 4-vinylbenzyl chloride (4VBC) and glycidyl methacrylate (GMA), was carried out. TNBS colorimetric assay, $^1$H‑NMR and ATR-FTIR confirmed covalent and tunable functionalization of collagen lysines. Depending on the type and extent of functionalization, controlled stability and thermal denaturation of triple helices were observed via circular dichroism (CD), whereby the hydrogen-bonding capability of introduced moieties was shown to play a major role. Full gel formation was observed following photo-activation of functionalized collagen solutions. The presence of a covalent network only slightly affected collagen triple helix conformation (as observed by WAXS and ATR-FTIR), confirming the structural organization of functionalized collagen precursors. Photo-activated hydrogels demonstrated an increased denaturation temperature (DSC) with respect to native collagen, suggesting that the formation of the covalent network successfully stabilized collagen triple helices. Moreover, biocompatibility and mechanical competence of obtained hydrogels were successfully demonstrated under physiologically-relevant conditions. These results demonstrate that this novel synthetic approach enabled the formation of biocompatible collagen systems with defined network architecture and programmable macroscopic properties, which can only partially be obtained with current synthetic methods.

---

[*] Corresponding author: David J. Wood (d.j.wood@leeds.ac.uk)

## 1. Introduction

Tissues, such as tendon or bone, are built upon complex macromolecular networks synthesized and organized by living cells.[1] The internal assembly of biological components defines tissue structure and shape. Tendons, for example, display large collagen fascicles at the macroscopic level, resulting from the arrangement of monomeric collagen into fibrils and fibres.[2] As a result, macroscopic tissues possess mechanical or other physical properties far superior to their constituents. The multi-scale hierarchical organization of biological components is therefore crucial to provide tissues with unique combination of compliance, mechanical strength and microstructure.[3]

A key challenge in the advent of *in vivo* tissue engineering strategies is the regeneration of damaged tissues via degradable tissue-like scaffolds,[4] providing a biomimetic interface to cells and exhibiting specific elasticity to the surrounding tissues.[5] Scaffolds providing cells with tissue-like hierarchical organization would be highly desirable in order to successfully restore structure, properties and functions in the neo-tissue. Collagen[6,7,8] offers great possibilities for biomimetic design of biomaterials, since it is the most abundant structural building block of connective tissues, conferring unique multi-scale organization, mechanical properties and biological functionality. Collagen has been widely applied for the design of vascular grafts,[9,10] fibrous materials for tissue engineering,[11] biomimetic scaffolds for alveolar- or tendon-like tissue regeneration,[12] and biocomposite matrices for hard tissue repair.[13] However, resulting collagen materials often exhibit restricted material properties, e.g. high swelling and poor elasticity, in physiological conditions, partially because collagen organization *in vivo* can only be partially reproduced *in vitro*. Consequently, the design of versatile collagen systems with defined protein conformation and enhanced macroscopic properties is still highly challenging. This challenge can only be overcome when understanding the material molecular organization, thereby establishing defined structure-property-function relationships.

From a molecular standpoint, the collagen molecule is based on three left-handed polyproline chains, each one containing the repeating unit Gly-*X-Y*, where *X* and *Y* are predominantly proline (Pro) and hydroxyproline (Hyp), respectively. The three chains are staggered to one another by one amino acid residue and are twisted together to form a right-handed triple helix (300 nm in length, 1.5 nm in diameter). *In vivo*, triple helices can aggregate to form collagen fibrils, fibres and fascicles, which are stabilized via covalent crosslinks.[14,15] Collagen fibrillogenesis can be induced *in vitro* by exposing monomeric collagen solutions to physiological conditions, resulting in viscoelastic gels at the macroscopic level.[16] The design of collagen mimetic peptides has also been proposed as an alternative strategy to recapitulate multi-scale organization of natural collagen.[17] However, despite formation of hierarchical triple helix assemblies, resulting thermal and mechanical stability is still not adequate for biomaterial applications.

Functionalization of side- or end-groups has been widely employed in linear biomacromolecules, e.g. gelatin,[18] and hyaluronic acid,[19] for the synthesis of amorphous hydrogels with varied molecular architectures and material properties. These methods have been applied to triple helical collagen.[20,21] In contrast to linear natural polymers, however, functionalization of collagen requires careful synthetic considerations, since the hierarchical collagen organization imposes constraints in terms of protein solubility, occurrence of functional groups available for chemical functionalization, and material biofunctionality. Collagen has been widely crosslinked with N-(3-Dimethylaminopropyl)-N′-ethylcarbodiimide hydrochloride (EDC),[8,22] glutaraldehyde (GTA)[23] and hexamethylene diisocyanate (HDI).[24] In the first case, zero-length covalent net-points are formed, so that no harmful and potentially cytotoxic molecules are introduced.[25] Due to the minimal net-point length, however, crosslinking of adjacent collagen fibrils is unlikely since terminal amino functions are too far to be bridged, resulting in non-varied mechanical properties.[23] Other than EDC, GTA and HDI involve the formation of oligomeric covalent net-points between distant polymer chains. Reaction of collagen with aldehydes or isocyanates in aqueous solution, has been reported to result in a cascade of non-controllable side reactions[20,23,24] and the formation of highly reactive and potentially toxic functional groups coupled to the polymer backbone.[26] To avoid undesirable side reactions, collagen was crosslinked with diimidoesters-dimethyl suberimidate (DMS), 3,3'-dithiobispropionimidate (DTBP) and acyl azide, proving to result in stable materials in physiological conditions, although material extensibility was found to be reduced.[27] Either dehydrothermal treatment or riboflavin-mediated photo-crosslinking are also applied as physical, benign crosslinking methods, although partial loss of native collagen structure and non-homogeneous crosslinking were observed.[28] Rather than direct covalent crosslinking, alternative approaches recently focused on the formation of injectable extracellular matrix-mimicking gels via synthetic collagen blends[29] as well as on the design of cell-populated matrices via derivatization with cinnamate[30] or acrylate[31] moieties. Here, while resulting mechanical properties may be enhanced, synthetic components, e.g. polymers or comonomers, were required to promote the formation of water-stable matrices, so that protein conformation, biofunctionality, and degradability may be affected.

The goal of this paper was to investigate whether covalent functionalization of collagen lysines could be accomplished with varied vinyl-bearing moieties, so that defined biomimetic systems with bespoke triple helical architecture could be established. By the synthesis of a photo-active collagen platform, injectable hydrogel networks were expected following *UV* irradiation. Material properties were therefore hypothesized to be controlled by the variation of the network architecture (dictated by the type of vinyl-bearing backbone, degree of collagen functionalization and concentration of functionalized collagen solution), whilst preserving the native collagen triple helical conformation. Following isolation in-house, type I collagen was reacted with either 4-vinylbenzyl chloride (4VBC) or glycidyl methacrylate (GMA). 4VBC was selected based on its backbone rigidity, hydrophobicity, and biocompatibility.[32] 4VBC-based systems were therefore expected to display reduced swellability and increased mechanical properties (i.e. compressability). In contrast, GMA was chosen to promote the formation of materials with enhanced elasticity; GMA has previously been employed for the design of flexible polymers,[33] non-cytotoxic protein-[34,35] and polysaccharide-based hydrogels.[36,37,38] Covalent functionalization with either 4VBC or GMA mainly occurs via the nucleophilic reaction of collagen $\varepsilon$-amino side groups with chlorine (4VBC) and epoxy (GMA) functions, respectively. Here, triethylamine[39] was used as catalyst, whilst tween-20 was applied as surfactant in order to mediate monomer miscibility in the aqueous phase. GMA- and 4VBC-functionalized collagens were characterized by 2,4,6-Trinitrobenzene sulfonic acid (TNBS) assay, Proton Nuclear Magnetic Resonance spectroscopy ($^1$H-NMR) and circular dichroism. Consequently, functionalized precursors were dissolved in solutions containing 2-Hydroxy-4′-(2-hydroxyethoxy)-2-methylpropiophenone (I2959), whereby I2959 was selected as a non-toxic photo-initiator.[40,41] Collagen-based systems were successfully formed following system photo-activation as proved by chemical (Attenuated Total Reflectance Fourier Transform Infrared spectroscopy, ATR-FTIR), structural (Wide Angle X-ray Scattering, WAXS) and thermo-mechanical analyses. Furthermore, evidence of biocompatibility was demonstrated via an indirect cytotoxicity assay. In this way, the triple helical network architecture was varied based on the selected, cell-friendly, functionalization step, so that programmable macroscopic properties were successfully achieved.

## 2. Experimental

### 2.1 Materials

Calf-skin type I collagen (CCS), glycidyl methacrylate (GMA), 4-vinylbenzyl chloride (4VBC), 2,4,6-trinitrobenzenesulfonic acid (TNBS) and Dulbecco's Phosphate Buffered Saline (PBS) were purchased from Sigma-Aldrich. Rat tails were supplied from the University of Leeds animal house. All the other chemicals were purchased from Sigma Aldrich.

### 2.2 In-house isolation of type I collagen from rat tail tendons

Type I collagen was isolated in-house via acidic treatment of rat tail tendons.[42] Briefly, frozen rat tails were thawed in distilled water. Individual tendons were pulled out of the tendon sheath, minced, and placed in 17.4 mM acetic acid solution at 4 °C in order to extract collagen. After three days extraction, the mixture was centrifuged at 20000 rpm for one hour. The supernatant was then freeze-dried in order to obtain type I collagen.

**2.3 Sodium dodecyl sulphate-polyacrylamide gel electrophoresis (SDS-page)**

In-house isolated collagen from rat rail tendons and commercially-available collagen from calf skin were dissolved in SDS sample buffer (160 mM Tris-HCl, pH 6.8, 2% SDS, 26% glycerol, 0.1% bromophenol blue) at 1 wt./vol.-% concentration and heated for 2 min at 90 °C. 10-30 $\mu$L of each sample solution were loaded onto 4% stacking gel wells and separated on 15% resolving gels (200 V, 45 min, room temperature). Protein bands were visualized after 60 min staining (0.1 wt.-% Comassie Blue, 12.5 vol.-% trichloroacetic acid) and 60 min treatment in water. The molecular weight of resulting bands was approximated by measuring the relative mobility of the standard protein molecular weight markers.

**2.4 Synthesis of functionalized collagen**

Type I collagen (0.25 wt.-%) was stirred in 10 mM hydrochloric acid solution at room temperature until a clear solution was obtained. Solution pH was neutralized to pH 7.4 to allow for collagen fibrillogenesis. Either 4VBC or GMA were added to the reaction mixture with 10-75 molar ratio with respect to collagen lysines (the collagen lysine content was determined via TNBS analysis). An equimolar amount of triethylamine (with respect to the added monomer) and 1 wt.-% of tween-20 (with respect to the collagen solution weight) were also applied. After 24 hours reaction, the mixture was precipitated in 10-15 volume excess of pure ethanol and stirred for two days. Ethanol-precipitated functionalized collagen was recovered by centrifugation and air-dried.

**2.5 Photo-activation and network formation**

4VBC-functionalized collagen was stirred at 4 °C in 10 mM HCl solution containing 1 wt.-% I2959. The resulting solution was cast onto a Petri dish, incubated in a vacuum desiccator to remove air-bubbles, followed by *UV* irradiation (Spectroline, 346 nm, 9 mW·cm$^{-2}$) for 30 min on each dish side. GMA-based collagen networks were prepared following the same protocol, except that the solution was prepared in PBS. Formed hydrogels were thoroughly washed in distilled water to remove unreacted compounds. Samples were air-dried following dehydration via an ascending series of ethanol-water mixtures (0-100% ethanol).

**2.6 Chemical characterization**

The degree of functionalization of collagen lysines was determined by 2,4,6-trinitrobenzenesulfonic acid (TNBS) colorimetric assay.[43] 11 mg of dry sample were mixed with 1 mL of 4 wt.-% $NaHCO_3$ (pH 8.5) and 1 mL of 0.5 wt.-% TNBS solution at 40 °C under mild shaking. After 4 hours reaction, 3 mL of 6 M HCl solution was added and the mixture was heated to 90 °C to dissolve any sample residuals. Solutions were cooled down and extracted three times with anhydrous ethyl ether to remove non-reacted TNBS species. All samples were read against a blank, prepared by the above procedure, except that the HCl solution was added before the addition of TNBS. The content of free amino groups and degree of functionalization ($F$) were calculated as follows:

$$\frac{moles(Lys)}{g(collagen)} = \frac{2 \times Abs(346\ nm) \times 0.02}{1.4 \times 10^4 \times b \times x}$$ (Equation 1)

$$F = 1 - \frac{moles(Lys)_{Funct.Collagen}}{moles(Lys)_{Collagen}}$$ (Equation 2)

where *Abs(346 nm)* is the absorbance value at 346 nm, $1.4 \cdot 10^4$ is the molar absorption coefficient for 2,4,6-trinitrophenyl lysine (in L/mol·cm$^{-1}$), $b$ is the cell path length (1 cm), $x$ is the sample weight, and *moles(Lys)$_{Funct.Collagen}$* and *moles(Lys)$_{Collagen}$* represent the lysine molar content in functionalized and native collagen, respectively.

Besides TNBS, collagen functionalization was also investigated by $^1$H-NMR (Bruker Avance spectrometer, 500 MHz) by dissolving 5 mg of dry samples in 1 mL deuterium oxide. ATR-FTIR was carried out on dry samples using a Perkin-Elmer Spectrum BX spotlight spectrometer with diamond ATR attachment. Scans were conducted from 4000 to 600 cm$^{-1}$ with 64 repetitions averaged for each spectrum. Resolution was 4 cm$^{-1}$ and interval scanning was 2 cm$^{-1}$.

**2.7 Investigation of collagen conformation**

Circular dichroism (CD) spectra of functionalized samples were acquired with a Jasco J-715 spectropolarimeter using 0.2 mg/mL solutions in 10 mM HCl. Sample solutions were collected in quartz cells of 1.0 mm path length, whereby CD spectra were obtained with 2 nm band width and 20 nm/min scanning speed. A spectrum of the 10 mM HCl control solution was subtracted from each sample spectrum. Temperature ramp measurements at 221 nm fixed wavelength were conducted from 20 to 60 °C with 20 °C/hour heating rate, so that denaturation temperature ($T_d$) was determined as the mid-point of thermal transition.

Protein conformation in photo-crosslinked collagen networks was investigated by WAXS. WAXS measurements were carried out on dry samples with a Bruker D8 Discover (40 kV, 30 mA, x-ray wavelength $\lambda = 0.154$ nm). The detector was set at a distance of 150 mm covering $2\theta$ from 5 to 40°. The collimator was 2.0 mm wide and the exposure time was 10 s per frame. Collected curves were subtracted from the background (no sample loaded) curve and fitted with polynomial functions ($R^2 > 0.93$). WAXS measurements were coupled with Differential Scanning Calorimetry (DSC) in order to investigate the thermal denaturation of collagen samples (TA Instruments Thermal Analysis 2000 System and 910 Differential Scanning Calorimeter cell base). DSC temperature scans were conducted with 10-200 °C temperature range and 10 °C·min$^{-1}$ heating rate. 10-15 mg sample weight was applied in each measurement and two scans were used for each sample formulation. The DSC cell was calibrated using indium with 20 °C·min$^{-1}$ heating rate under 50 cm$^3$·min$^{-1}$ nitrogen atmosphere.

**2.8 Compression tests**

Water-equilibrated hydrogel discs (ø 0.8 cm) were compressed at room temperature with a compression rate of 3 mm·min$^{-1}$ (Instron 5544 UTM). A 500 N load cell was operated up to sample break. The maximal compressive stress ($\sigma_{max}$) and compression at break ($\varepsilon_b$) were recorded, so that the compressive modulus ($E$) was calculated by fitting the linear region of the stress-strain curve.

**2.9 Extract cytotoxicity assay**

An extract cytotoxicity assay was conducted with L929 mouse fibroblasts and $\gamma$-sterilized collagen hydrogels following European norm standards (EN DIN ISO standard 10993-5). 0.1 mg of hydrogel sample was incubated in 1 mL cell culture medium (Dulbecco's Modified Eagle Medium, DMEM) at 37 °C. After 72-hour incubation, the sample extract was applied to 80% confluent L929 cells. Dimethyl sulfoxide (DMSO) was used as negative control while cell culture medium was applied as positive control. Cell morphology was monitored using a transmitted light microscope in phase contrast mode.

**3. Results and discussion**

Collagen was functionalized with varied vinyl-bearing moieties aiming at the formation of water-stable biomimetic systems with varied triple helical network architecture. Collagen was isolated in-house from rat tail tendons, resulting in a defined, readily-available protein building block for tuneable, covalent functionalization. By reaction with either 4-vinylbezyl chloride (4VBC) as rigid monomer, or glycidyl methacrylated (GMA) as flexible monomer, hydrogel networks were successfully obtained following *UV* irradiation (Scheme 1).

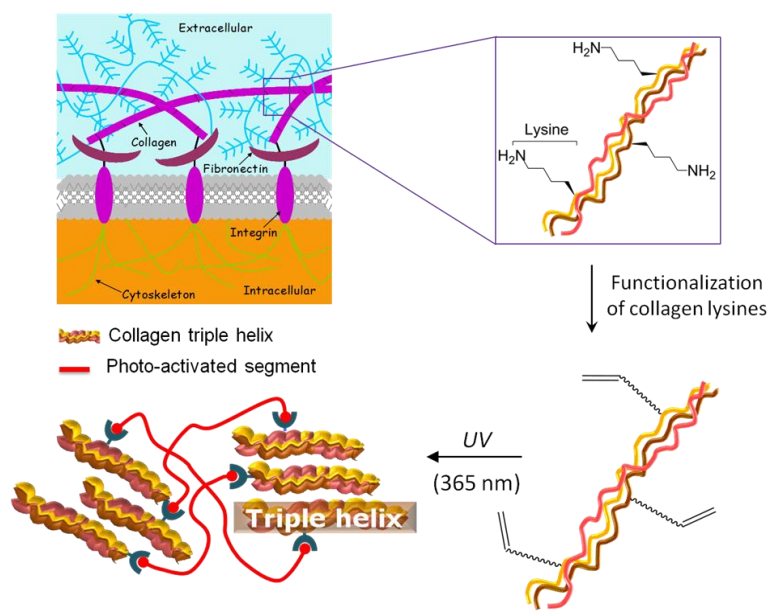

**Scheme 1.** Design of triple-helical collagen-based hydrogels. Collagen is isolated in-house from rat tail tendons and selectively functionalized with photo-active compounds of varied backbone rigidity. *UV* irradiation in the presence of a cyto-compatible photo-initiator successfully leads to the formation of photo-crosslinked hydrogels. In this way, cyto-compatible, defined network architectures are accomplished with controlled material stability.

Samples of functionalized collagen are identified as AAA-BBBY, whereby AAA indicates the type of collagen used in the synthesis, either in-house isolated type I collagen from rat tails (CRT), or commercially-available type I collagen from calf skin (CCS). BBB designates the system of functionalized collagen, i.e. either GMA- or 4VBC-based system. Y represents the monomer/lysine feed molar ratio of either GMA or 4VBC. In the case of collagen networks, samples are coded as AAAX-BBBY$^*$, whereby '*' denotes a collagen network, X identifies the solution concentration of functionalized collagen, while AAA, BBB and Y have the same meaning as previously-stated.

## 3.1 Isolation of type I collagen from rat tail tendons

In-house isolated collagen was characterized by SDS page as for its molecular weight distribution (Figure 1). Following electrophoretic denaturation, a mixture of monomeric α-chains (~100 kDa), dimeric β-components (~200 kDa), i.e. two covalently crosslinked α-chains ($[α_1(I)]_2$), and trimeric γ-components (~300 kDa), consisting of three covalently crosslinked α-chains ($[α_1(I)]_2[α_2(I)]$) were observed.[44] Each of these species was identified in the electrophoretic patterns of in-house isolated collagen (CRT-1–3) as well as in the case of commercially-available collagen type I from calf skin (CCS-1–3). Furthermore, a weakly-stained band at around 50 kDa was observed in CRT and in CCS (even if at a lower extent) patterns, likely hinting at a degraded collagen species[45] formed following sample denaturation as a consequence of the heating step in the preparation of the SDS gel. Overall, CRT accurately displayed electrophoretic reference bands of collagen and was therefore applied as protein backbone for subsequent covalent functionalization.

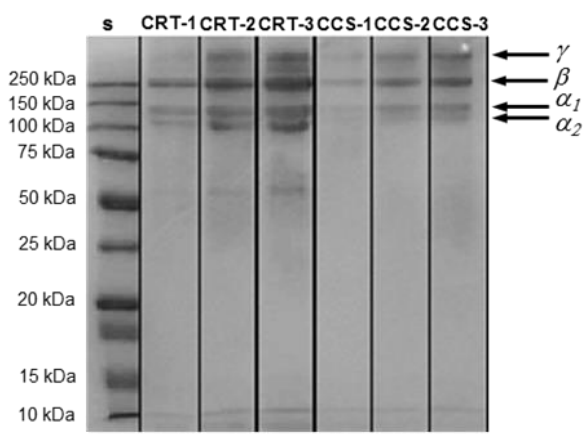

**Figure 1.** SDS-page analysis of standard (S), in-house isolated type I collagen from rat-tail (CRT-1–3), and commercially-available type I collagen from calf skin (CCS-1–3).

### 3.2 Chemical functionalization of type I collagen

Collagen functionalization with either 4VBC or GMA occurs via nucleophilic addition of collagen side-terminations to chlorine and epoxy functionalities, respectively. Potential functional groups involved in the reaction include the $ε$-amino functions of lysine and hydroxylysine, hydroxyl functions of serine, threonine and tyrosine, as well as thiol groups of cysteine. Of these functional groups, $ε$-amino functions of lysines are known to be highly reactive species and will therefore be predominant, resulting in a chemo-selective reaction (Scheme 2).

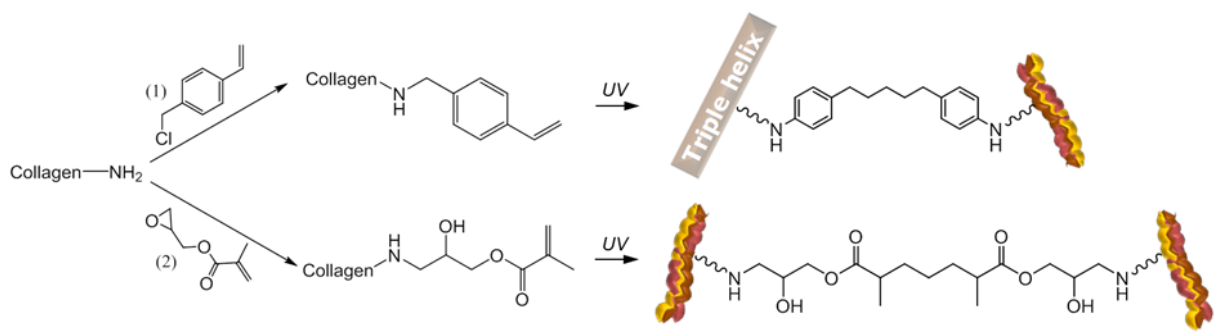

**Scheme 2.** Synthesis of functionalized and photo-activated systems. Collagen is reacted with 4-vinylbenzyl chloride (4VBC, 1) and glycidyl methacrylate (GMA, 2), respectively. Following *UV* irradiation of collagen precursors solutions, covalent net-points are introduced between collagen triple helices (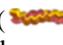), so that photo-crosslinked networks with varied triple-helical architecture are successfully accomplished.

For the reaction to take place, a non-acidic solution pH is crucial in order to avoid amino group protonation and to enhance lysine reactivity. For these reasons, acidic collagen solution was neutralized to pH 7.4. These pH conditions are normally applied to induce collagen fibrillogenesis *in vitro*, whereby a less transparent, cloudy mixture was observed. This optical observation was confirmed by a rapid increase in solution optical density (Figure 1, Supp. Inf.), thereby suggesting the reconstitution of collagen triple helices into native-type fibrils.[8] Consequently, a further proof of in-house collagen type I isolation was obtained.

Following reaction of collagen with either 4VBC or GMA, TNBS colorimetric assay was applied to resulting products in order to assess the molar content of free, non-reacted *ε*-amino groups and quantify the degree of collagen functionalization. Type I collagen generally presents $33 \cdot 10^{-5}$ moles(*Lys*)·g$^{-1}$;[22-24] this value was similar to the lysine content observed in in-house isolated CRT. On the other hand, a much lower lysine content (~ $18 \cdot 10^{-5}$ moles·g$^{-1}$) was observed in CCS, potentially ascribed to the different tissue source, being tendon in CRT and skin in CCS. Given the higher lysine content in CRT compared to CCS, CRT was expected to result in a wider range of functionalization and was therefore preferred as readily available starting building block for the formation of collagen hydrogels. The reaction with either 4VBC as rigid monomer, or GMA as flexible monomer, was conducted with varied monomer-to-lysine molar ratios. Reaction products displayed a lowered molar content of free, non-reacted lysines (372±17 → 145±1 $\mu$moles·g$^{-1}$, Table 1), as observed by a decreased 346 nm-absorbance peak in functionalized, in contrast to native collagen (Figure 2, left). Consequently, covalent functionalization, rather than simple physical blend, of collagen with both monomers was confirmed (Table 1). The degree of functionalization (*F*) could be adjusted between 0-61 mol.-% in samples CRT-GMA (Figure 2, right), while a lower range of functionalization was observed in samples CRT-4VBC (*F*: 0-35 mol.-%). The monomer-dependent *F* profiles are likely to be explained based on the different miscibility and reactivity of 4VBC and GMA in aqueous collagen solutions.

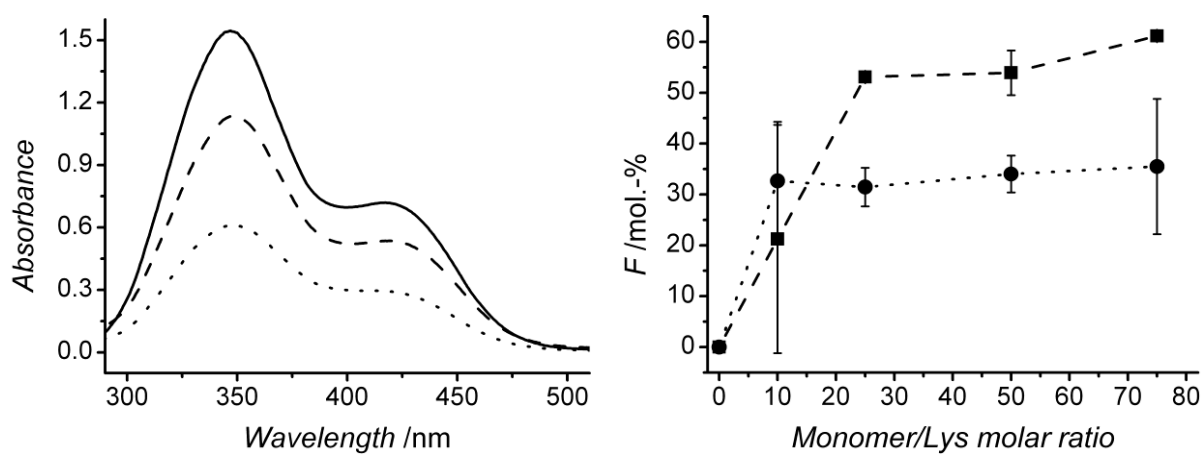

**Figure 2.** Left: Absorbance curves resulting from TNBS assay on in-house isolated (—) and functionalized (CRT-4VBC50 (--); CRT-GMA75 (··)) type I collagens. Right: degree of collagen functionalization (*F*) with either 4VBC (··●··) or GMA (-■-) following reaction with varied monomer-*Lys* molar ratios.

In order to further explore the covalent functionalization of collagen lysines, $^1$H-NMR spectra of functionalized and native collagens were recorded. Figure 3 displays exemplary spectra of CRT (A), CRT-4VBC blend (B, 50 4VBC/Lys molar excess) and CRT-4VBC25 (C). Here, geminal protons of 4VBC (at 5.2-6.7 ppm$^{32}$) were successfully identified in the functionalized sample in contrast to the CRT control. At the same time, the CRT-4VBC blended mixture was also analysed as an additional control; here, the $^1$H-NMR spectrum mainly displays 4VBC-related peaks, while the presence of CRT was not detected, likely related to the excess of 4VBC with respect to the collagen lysines. With this investigation, it was therefore demonstrated that $^1$H-NMR spectrum of the blend CRT-4VBC was very different from ¹H-NMR spectra of both CRT and CRT-4VBC25; furthermore, it was also demonstrated that ¹H-NMR spectrum of CRT-4VBC25 differed from the one of CRT by only the presence of 4VBC germinal peaks. Consequently, further evidence of covalent functionalization rather than simple blend formation in reacted samples was provided, in agreement with TNBS assay results. Moreover, it was proved that the precipitation in ethanol successfully enabled the purification of reacted products from non-reacted species (e.g. monomers and surfactant).

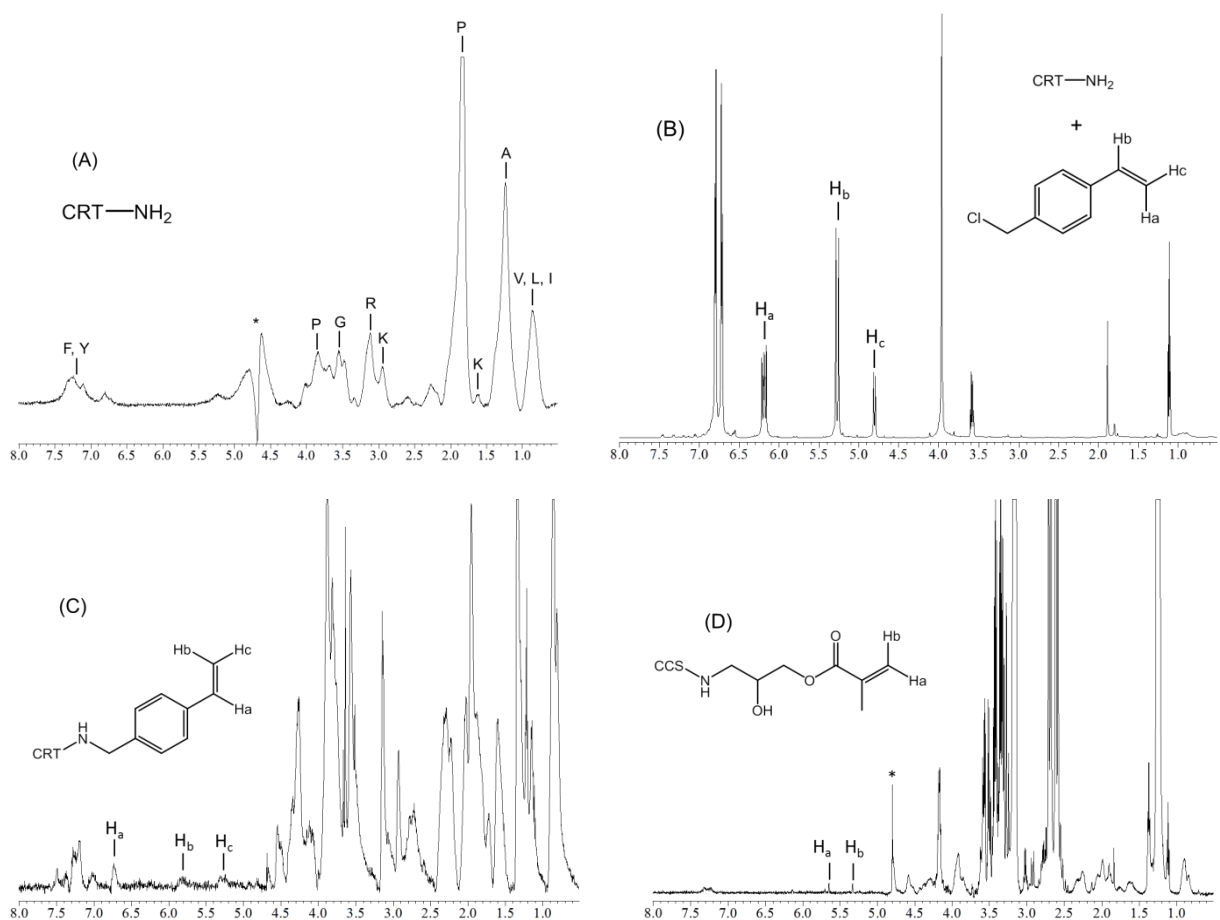

**Figure 3.** ¹H-NMR spectra of (A) CRT, (B) CRT-4VBC blend (50 4VBC-Lys molar excess), (C) CRT-4VBC25, (D) CCS-GMA90. ¹H-NMR peaks of vinyl geminal protons are depicted in the region 5.15-6.68 ppm, confirming TNBS results of collagen functionalization.

In order to investigate the versatility of the reaction, CCS derived type I collagen was applied as an alternative protein backbone for the reaction with GMA. Here, a degree of functionalization of 26 ± 5 mol.-% was obtained, which was supported by the ¹H-NMR spectrum (Figure 3, D) of the resulting product (geminal proton peaks at 5.3-5.6 ppm[31]). The decreased *F* in CCS-compared to CRT-based products is probably due to the lower concentration of CCS (0.1 wt.-%) used in the reaction compared to the one of CRT (0.25 wt.-%).

Overall, the presence of triethylamine as catalyst and tween-20 as surfactant was crucial to increase the reaction yield, while the combination of TNBS and $^1$H-NMR was necessary to confirm covalent attachment of vinyl moieties rather than physical monomer incorporation. In this way, it was possible to functionalize collagen in a chemo-selective and tuneable manner so that *F* was successfully adjusted based on the monomer type and monomer feed ratio.

**3.3 Investigation of protein conformation in functionalized collagen**

A major challenge in the formation of collagen-based materials is the application of synthetic methods which, on the one hand, enable controlled material stability in physiological conditions and, on the other hand, preserve native protein conformation. Single polyproline chains of collagen are stabilized into a triple helix structure via hydrogen bonds oriented perpendicularly to the triple helix axis, resulting in an optically active protein.[46] Far-*UV* circular dichroism (CD) spectroscopy was therefore applied to characterize collagen conformation before and after covalent functionalization. This was then coupled with temperature-ramp measurements in order to investigate the effect of functionalization on the thermal stability of collagen triple helices.

Far-*UV* CD measurements were conducted on sample solutions in slightly acidic environments, whereby in-house isolated CRT was compared with CCS as well as CRT-4VBC and CRT-GMA samples (Figure 4). A spectrum of gelatin, as partially-denatured collagen, was also recorded as an additional control. Based on its polyproline II-like helical assembly, the collagen CD spectrum displays a positive maximum absorption band at 210-230 nm, as related to the triple helix conformation,[47] and a negative minimum absorption band around 190 nm, showing the random coil conformation.[48]

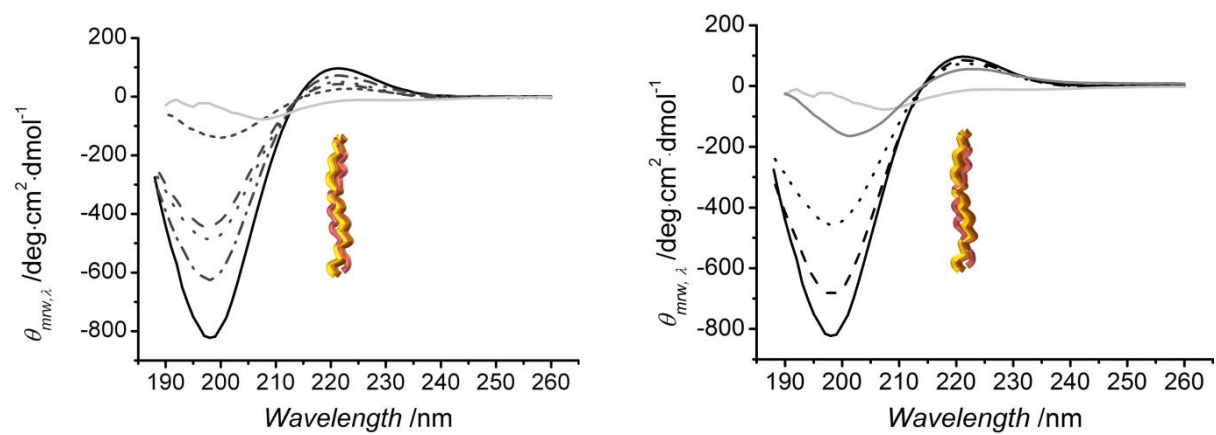

**Figure 4.** Left: Far-UV CD spectra of samples CRT-4VBC. (–·–): CRT-4VBC10, (– –): CRT-4VBC25, (--): CRT-4VBC50, (···): CRT-4VBC75. Right: Far-UV CD spectra of CRT-GMA samples. (– –) CRT-GMA25, (···) CRT-GMA75. Controls of collagen and gelatin are provided: (–) CRT, (–) CCS, (–) gelatin.

Both peaks were observed in the case of native samples CCS and CRT, confirming the typical conformation of collagen in the in-house isolated CRT. In contrast to gelatin, the positive absorbance band was observed in samples CRT-4VBC and CRT-GMA, suggesting that triple helix conformation was preserved following covalent functionalization of lysines. This finding is in agreement with CD observations on collagen after reaction with methacrylic moieties.[31,49] Whilst in the case of samples CRT-GMA where the triple helix structure was completely preserved, the positive absorbance band intensity appeared to be decreased in the spectra of samples CRT-4VBC. This suggests that the introduction of aromatic, bulky groups may influence the collagen conformation in resulting functionalized samples. This is an interesting finding and may be explained in terms of varied hydrogen bonding capabilities and steric effects of incorporated vinyl-bearing backbones. Lysine terminations are known to stabilize collagen triple helices via hydrogen bonds with other polyproline chains;[50] on the other hand, aromatic residues are expected to destabilize the trimer due to their inability in forming hydrogen bonds due to the absence of acceptor/donor hydrogen groups. In the case of GMA-functionalized collagen, new hydroxyl as well as methacrylic carbonyl groups will be formed following nucleophilic reaction of collagen lysines with GMA epoxide ring (Scheme 2). Both functionalities are supposed to promote hydrogen bonds, thereby mediating triple helix stabilization. In contrast, lysine functionalization with 4VBC will result in the coupling of vinyl benzene residues (Scheme 2), which are unlikely to act as hydrogen bond donor/acceptor species.

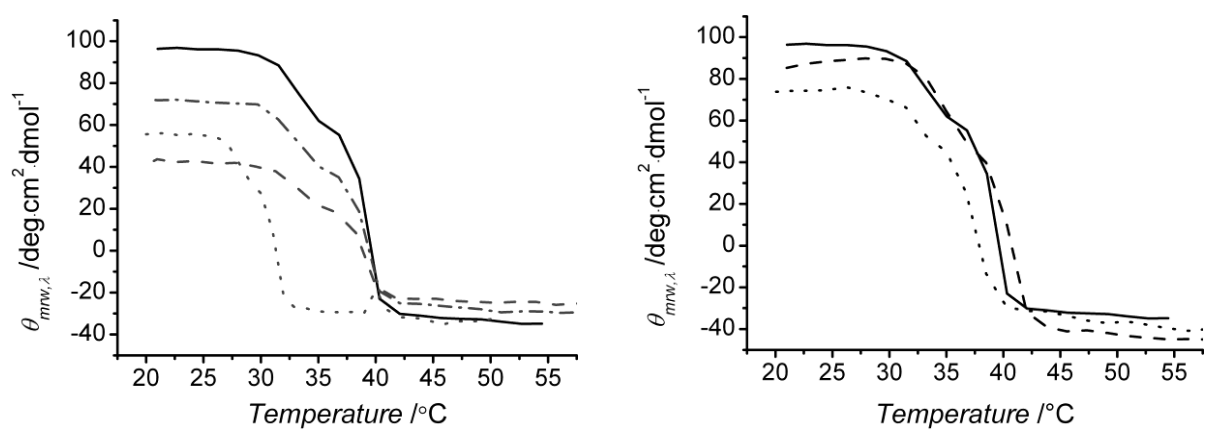

**Figure 5.** Temperature-ramp CD spectra of CRT-4VBC (left) and CRT-GMA (right) samples at 221 nm fixed wavelength. Left: (–): CRT, (–·–) CRT-4VBC10, (--) CRT-4VBC50, (···) CRT-4VBC75. Right: (–): CRT, (– –) CRT-GMA25, (···) CRT-GMA75.

This, together with considering the relative bulkiness of the introduced aromatic residues, is likely to explain the partially-reduced peak intensity of collagen triple helices in CD spectra of 4VBC-functionalized samples.

In addition to far-*UV* CD, temperature ramp measurements were also carried out in order to investigate thermal denaturation of functionalized collagen. As described in Figure 5 (right), 221 nm-molar ellipticity was observed to decrease in native collagen, reflecting heating-related triple helix denaturation ($T_d$ ~ 39 °C). A similar mean residue ellipticity profile was also observed in GMA-functionalized collagen, resulting in comparable denaturation temperatures ($T_d$ ~ 35–39 °C, Table 1). Consequently, further evidence was obtained that the coupling of GMA moieties onto collagen lysines did not affect structural/thermal properties of collagen solutions. Likewise, samples CRT-4VBC showed a comparable decrease in 221 nm-mean residue ellipticity upon heating ($T_d \sim$ 30–37 °C, Table 1).

**Table 1.** Chemical and structural properties of functionalized collagens. *Lys*: free lysine content; $V_c$: total vinyl content; *Y*: yield of functionalization reaction; $T_d$: triple helix denaturation temperature as determined by temperature-ramp CD at 221 nm-fixed wavelength.

| Sample ID | $Y$ /wt.-% | $Lys$ /μmol·g$^{-1}$ | $V_c$ /μmol·g$^{-1}$ | $T_d$ /°C |
|---|---|---|---|---|
| CRT–4VBC10 | 84 | 251 ± 43 | 122 ± 43 | 35 |
| CRT–4VBC25 | 86 | 255 ± 14 | 117 ± 14 | 37 |
| CRT–4VBC50 | 83 | 246 ± 13 | 127 ± 13 | n.a. |
| CRT–4VBC75 | n.a. | 240 ± 49 | 132 ± 49 | 30 |
| CRT–GMA10 | 77 | 293 ± 83 | 79 ± 83 | 37 |
| CRT–GMA25 | 85 | 175 ± 2 | 198 ± 2 | 39 |
| CRT–GMA50 | 81 | 172 ± 16 | 201 ± 16 | 37 |
| CRT–GMA75 | n.a. | 145 ± 1 | 228 ± 1 | 35 |

This is an interesting finding, since samples functionalized with aromatic residues displayed reduced triple helix-related peak intensity, so that a much decreased thermal stability may be expected. However, although aromatic residues are not supposed to form triple helix-stabilizing hydrogen bonds, they are well-known to mediate other secondary interactions, i.e. π-π stacking or hydrophobic interactions.[32] It is therefore likely that these physical interactions account for the comparable thermal stability of 4VBC-based solutions in comparison with native collagen solutions.

### 3.4 Photo-activation of functionalized collagens and network formation

Once the synthesis and characterization of functionalized collagen was explored, the attention moved to the formation of water-stable hydrogels via photo-activation of functionalized collagen. CRT-4VBC sample solutions were prepared in I2959-10 mM HCl solutions, while samples CRT-GMA were dissolved in I2959-PBS solution, in order to explore the formation of an injectable material under physiologically-relevant conditions. Remarkably, *UV* irradiation of both types of functionalized collagen solutions proved to result in water-stable gels following 30 min irradiation, among the whole set of sample compositions. In contrast, no gel formation was observed in the case of native collagen solution (in the presence of I2959) as well as in the case of collagen-monomer blends. These findings suggest that resulting hydrogels could only be obtained following photo-activation of collagen-coupled vinyl moieties, thereby leading to the formation of a covalent network.

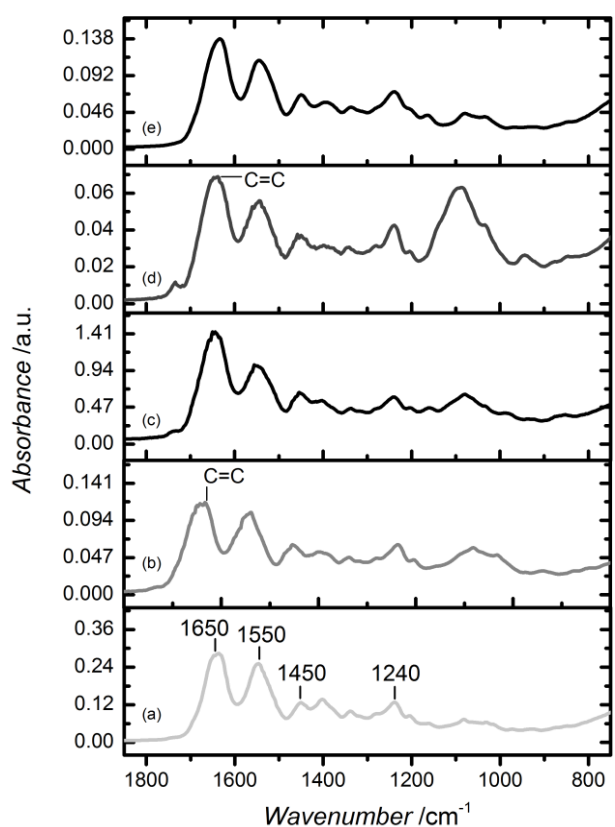

**Figure 6.** ATR-FTIR spectra of collagen-based samples. (a): CRT, (b): CRT-GMA50, (c): CRT1-GMA50[*], (d): CRT-4VBC10, (e): CRT1-4VBC10[*]. Main collagen bands are displayed in CRT spectrum. In contrast to CRT, both samples CRT-GMA50 and CRT-4VBC reveal additional peaks at 1630-1640 $cm^{-1}$, which are suppressed in the photo-activated systems spectra. This investigation gives supporting evidence that network formation is successfully obtained at the molecular level, while preserving the triple helical collagen conformation.

Following optical observations of material preparation, it was important to confirm the formation of a covalent network following *UV* irradiation. This was accomplished by ATR-FTIR spectroscopy on dry 4VBC- and GMA-based networks, as well as on dry native and functionalized precursors (Figure 6). The presence of 4VBC and GMA was expected to be associated with an absorption band at 1630-1640 cm$^{-1}$ in resulting ATR-FTIR spectra, as this band identifies the vibration of C=C double bonds.[34,51] These peaks were observed in the case of functionalized collagens, confirming TNBS and $^1$H-NMR results, while they were suppressed in the case of photo-crosslinked and native collagen. In light of this finding, it was concluded that the photo-crosslinking reaction successfully occurred during *UV* irradiation.

**3.5 Structural organization in photo-activated systems**

Besides investigation on the chemical structure of *UV*-irradiated materials, ATR-FTIR was employed to elucidate collagen conformation in the network, as previously addressed via solution-based CD spectroscopy in functionalized precursors. Triple helix collagen conformation is normally associated with three main amide bands in ATR-FTIR collagen spectrum, i.e. amide I at 1650 cm$^{-1}$, resulting from the stretching vibrations of peptide C=O groups; amide II absorbance at 1550 cm$^{-1}$, deriving from N–H bending and C-N stretching vibrations; and amide III band centred at 1240 cm$^{-1}$, assigned to the C-N stretching and N-H bending vibrations from amide linkages, as well as wagging vibrations of CH$_2$ groups in the glycine backbone and proline side chains.[52] The positions of these amide bands are observed in the spectrum of CRT and are maintained in the spectra of functionalized and photo-activated samples (Figure 6). Furthermore, the FTIR absorption ratio of amide III to 1450 cm$^{-1}$ band was determined to be close to unity ($A_{III}/A_{1450}$ ~ 0.93–1.01) among the different samples. This result provides clear evidence that the collagen conformation was not altered following hydrogel preparation, from covalent functionalization to network formation. This is in agreement with previous findings, as both sample precursors CRT-GMA50 and CRT-4VBC10 showed nearly preserved triple helical conformation in far-*UV* CD analysis (Figure 4).

In addition to FTIR, WAXS was applied to further elucidate the effect of molecular network architecture on the triple helical conformation. Consequently, non-soluble collagen networks were analysed in the dry state (based on the fact that there is only a minimal reported difference in collagen molecular organization with respect to the wet state[53]). WAXS has been applied to collagen samples in order to get information about the packing features of collagen in terms of distances between collagen molecules in the lateral plane of the collagen fibril (intermolecular lateral packing) and distances between amino acids along the polypeptide chain (helical rise per residue).[53,54,55] Figure 7 (left) displays WAXS spectra of linear intensity vs. scattering vector resulting from samples CRT, CRT1-4VBC25* and CRT1-GMA50*. As expected, WAXS spectrum of CRT displays three main collagen peaks, identifying the intermolecular lateral packing of collagen molecules ($d \sim 1.1$ nm, $2\Theta \sim 8°$), the isotropic amorphous region ($d \sim 0.5$ nm, $2\Theta \sim 20°$) and the axial periodicity ($d \sim 0.29$ nm, $2\Theta \sim 31°$) of polypeptide subunits *(Gly-X-Y)* along a single collagen chain. Besides native collagen, both samples CRT1-4VBC25* and CRT1-GMA50* highlight a similar WAXS spectrum, so that the 1.1 nm peak corresponding to the triple helix packing is still present following network formation, regardless of the network architecture. Whilst the peak positions are completely maintained in the spectrum of sample CRT1-GMA50*, slight peak shifts are present in the spectrum of sample CRT1-4VBC25* (with respect to CRT), indicating an alteration in the native collagen packing features. The observations deriving from WAXS analysis on covalent networks are in agreement with previous CD results obtained on functionalized precursors, supporting the fact that the slight change in collagen conformation in CRT1-4VBC25* is mainly related to the incorporation of aromatic moieties rather than to the presence of a covalent network. At the same time, native collagen packing is completely preserved in the case of GMA-based systems, again providing further evidence that functional groups present on GMA backbone can mediate triple-helix stabilizing hydrogen bonds.

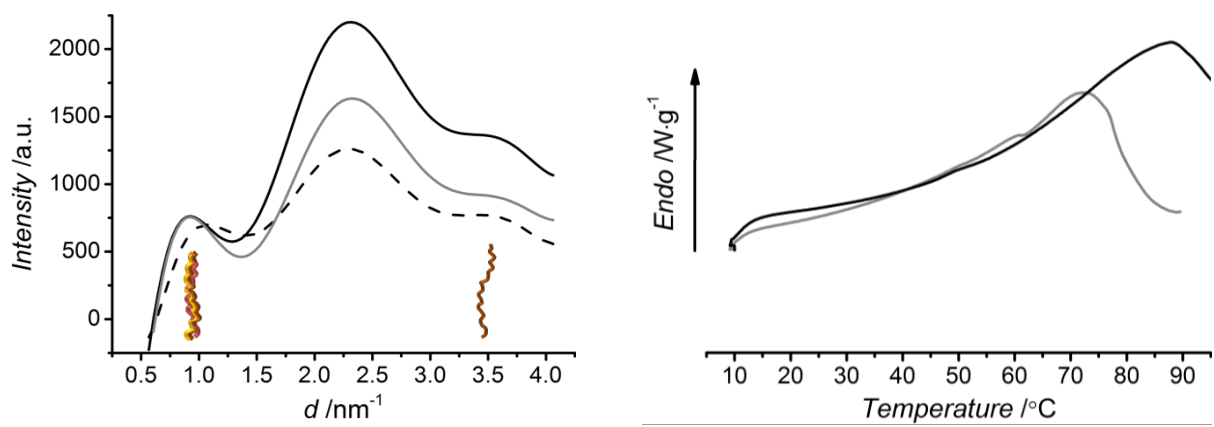

**Figure 7.** WAXS spectra (left) and DSC thermograms (right) of CRT (gray), CRT1-GMA50* (solid black) and CRT1-4VBC25* (dashed black) samples. Triple helical organization is observed in WAXS spectra of crosslinked collagen, whose thermal stability (DSC) is increased compared to native CRT, due to the presence of covalent net-points.

### 3.6 Thermo-mechanical analysis and cytotoxicity study in photo-activated systems

Following elucidation of the protein organization, dry collagen networks were equilibrated in aqueous solution so that resulting thermal properties were analyzed via DSC analysis. Figure 7 (right) depicts exemplary DSC thermograms of native and photo-activated systems. One endothermic transition is observed in both spectra in the range of 70-90 °C. Given that both materials were proved to be based on triple helical collagen architecture, the endothermic peak is likely to identify the melting denaturation of triple helices.[52]

**Table 2.** Shrinking temperature ($T_s$) in native and photo-activated collagen as quantified via DSC.

| Sample ID | CRT | CRT1-GMA10* | CRT1-GMA25* | CRT0.7-GMA25* | CRT1-GMA50* | CRT0.7-GMA50* | CRT1-GMA75* |
|---|---|---|---|---|---|---|---|
| $T_s$ /°C | 67±7 | 64±2 | 79±3 | 79±2 | 105±1 | 88±11 | 81±9 |

DSC has been widely applied to characterize the shrinking temperature ($T_s$) of crosslinked collagen-based samples,[56] at which temperature unfolding of collagen triple helices into randomly-coiled chains occur, resulting in nearly 80% material length reduction.[22-24] $T_s$ is therefore expected to be highly affected by the formation of a covalent network. Table 2 describes hydrogel $T_s$ values as obtained by DSC. $T_s$ values of CRT are determined in the same temperature range of other collagen materials,[22] suggesting a triple helical denaturation at around 70 °C. At the same time, $T_s$ is found to be increased (70 → 105 °C) following incorporation of triple helices into a network architecture. This finding confirms that covalent net-points are successfully introduced between collagen triple helices (Scheme 2), thereby enhancing hydrogel thermal properties.[52] Interestingly, hydrogels deriving from functionalized systems with increased degree of functionalization displayed an increase of $T_s$ values; this suggests that it is possible to adjust hydrogel thermal stability based on the molecular architecture of resulting covalent network. It should be noted that resulting hydrogels revealed higher denaturation temperatures compared to EDC,[22] glutaraldehyde-,[23] and hexamethylene diisocyanate-crosslinked[24] collagen. Thus, the presented photo-activation of functionalized collagen superiorly stabilizes collagen molecules in comparison with current crosslinking methods.

Other than thermal analysis, the mechanical properties of hydrogels CRT1-4VBC50* and CRT1-GMA50* were exemplarily determined by wet-state measurements to obtain evidence of the mechanical competence of formed materials in physiologically-relevant conditions. Samples described *J*-shaped stress-compression curves (Figure 8), similar to the case of native tissues.[11] Furthermore, shape recovery was observed in both hydrogels following load removal up to nearly 50% compression, confirming that the established covalent network successfully resulted in the formation of an elastic material. Compressive moduli ($E$) were measured in the kPa range, while compression at break ($\varepsilon_b$) did not exceed 70% compression (CRT1-4VBC50*: $E$ ~ 114 kPa, $\varepsilon_b$ ~ 35 %; CRT1-GMA50*: $E$ ~ 62 kPa; $\varepsilon_b$: 70 %). Most importantly, the selected network architectures were found to directly impact the mechanical properties of the hydrogels; here the incorporation of rigid aromatic moieties (Scheme 2) led to the formation of stiff, less elastic materials; at the same time, collagen functionalization with aliphatic, flexible backbones (Scheme 2) provided hydrogels with increased compressability and decreased compressive modulus. Therefore, it was possible to govern the macroscopic hydrogel properties depending on the specific network architecture, as observed in the case of linear, biopolymer-based systems.[5]

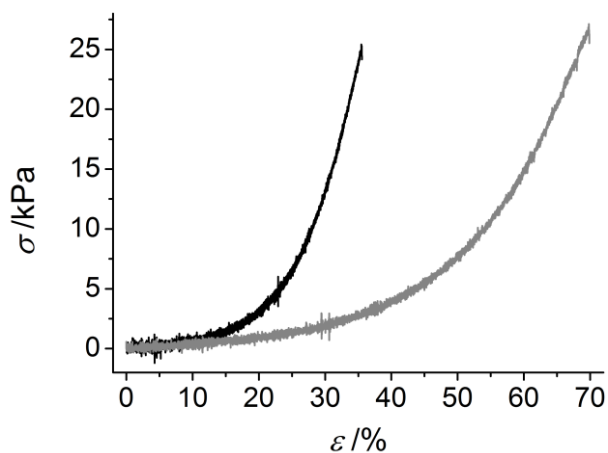

**Figure 8.** Exemplary stress-compression curves of CRT1-4VBC50[*](—) and CRT1-GMA50[*](—). Compressive moduli were obtained by fitting the linear region of the curves.

Following physical characterization of the hydrogels, the potential material performance in biological environment was investigated by an extract cytotoxicity assay following European guidelines. Gamma-sterilized samples CRT-GMA50[*] were exemplarily incubated in cell culture medium, so that the extracted supernatant was used for cell culture with L929 mouse fibroblasts. Cell response to hydrogel extracts was investigated after 48 hours cell culture by qualitative cell morphology observations (Figure 9). L929 mouse fibroblasts exhibited a spread-like morphology when cultured in either hydrogel extract or cell culture medium, suggesting that the material extracts did not induce any negative effect on cell proliferation. These observations provide supporting evidence that no non-reacted, potentially toxic compounds are present in the resulting materials. In light of these exemplary cell culture tests, future steps will focus on a systematic investigation involving specific cells in contact with systematically-varied network hydrogel architectures, in order to explore how changes at the molecular and macroscopic material level influence cell response.

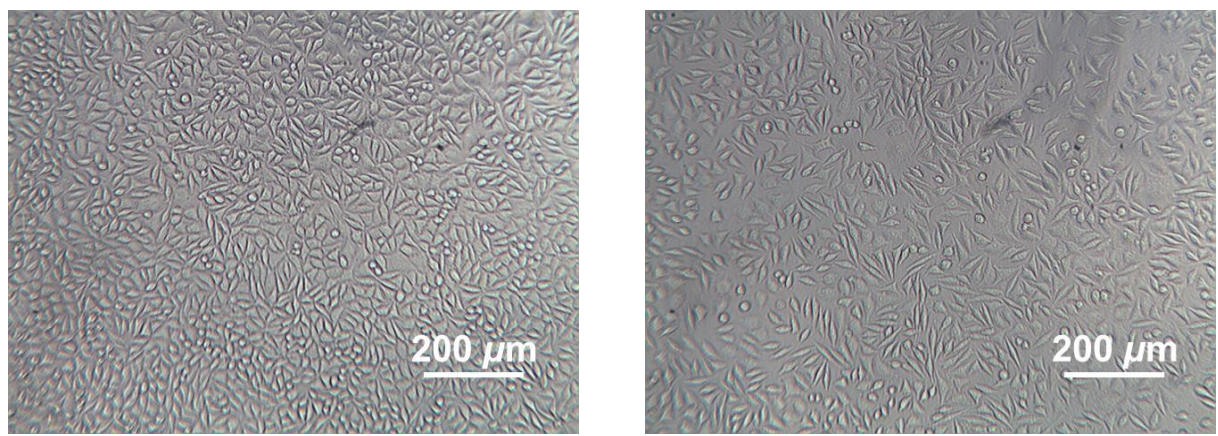

**Figure 9.** L929 cell morphology after 48 hours cell culture in cell culture medium (left) and 72-hours sample extract (CRT1-GMA50[*], right).

## 4. Conclusion

Photo-active collagen-based systems were synthesized as a platform for the establishment of defined biomimetic materials. Type I collagen was covalently functionalized with two monomers of varied flexibility, either 4VBC or GMA. The effect of monomer feed ratio on the extent of collagen functionalization and structural organization was investigated. Reaction with GMA provided a covalent and tunable degree of functionalization (as supported by TNBS, $^1$H-NMR, and ATR- FTIR) of collagen lysines ($F$: 0-61 mol.-%), while a lowered range of functionalization was found in the case of 4VBC-functionalized samples ($F$: 0-35 mol.-%), likely due to the reduced miscibility and reactivity of 4VBC in aqueous collagen solutions. Resulting photo-active systems displayed controlled triple helix conformation. Collagen molecule organization and triple helix thermal denaturation were preserved in GMA-functionalized products, suggesting that incorporation of GMA moieties help in mediating triple helix-stabilizing hydrogen bonding. Conversely, 4VBC-functionalized collagens revealed a decreased intensity of CD triple helix band intensity, likely due to the fact that introduced aromatic, bulky groups proved to hinder the formation of hydrogen bonds among protein single chains. Here, the thermal stability of 4VBC-functionalized system was still comparable to that of native collagen, suggesting that additional secondary interactions, e.g. π-π hydrophobic interactions, among aromatic residues, are likely to be established among functionalized collagen molecules. Photo-activation of functionalized systems resulted in the formation of water-stable hydrogels, as confirmed by the presence of a covalent network at the molecular level (ATR-FTIR). Structural material organization displayed preserved triple helical conformation (ATR-FTIR, WAXS), suggesting that the photo-activation step maintained the same protein organization as observed in the case of functionalized systems. Resulting hydrogels revealed an increased triple helix thermal stability, whereby $T_s$ was found to be affected by the degree of collagen functionalization (DSC). At the same time, variations in triple helical network architecture were directly related to changes in macroscopic mechanical properties, so that the backbone rigidity of introduced moieties was key to obtaining tuneable compressability and compressive modulus. Also in light of the observed hydrogel cyto-compatibility, next steps will focus on a thorough investigation of the macroscopic properties and material biofunctionality of these covalently-crosslinked collagen systems.

**Acknowledgements**

This work was funded through WELMEC, a Centre of Excellence in Medical Engineering funded by the Wellcome Trust and EPSRC, under grant number WT 088908/Z/09/Z. The authors wish to thank Dr. S. Brookes, Dr. S. Maude and Dr. J. Fisher, G. Nasir Khan, and J. Hudson for their kind assistance with SDS-page, NMR, CD and SEM analyses, respectively.